\documentclass[11pt,twoside]{article}
\usepackage{asp2004}
\usepackage{psfig}
\usepackage{epsf}
\usepackage{graphics}
\usepackage{lscape}
\newcommand{\kms}{\hbox{\,km\,s$^{-1}$}}

\begin{document}
\title{The outer ejecta of $\eta$ Carinae }
\author{K.\ Weis}
\affil{Astronomisches Institut, Ruhr-Universit\"at Bochum,
  Universit\"atsstr.\,150, 44780 Bochum, Germany}

\begin{abstract}
The nebula around $\eta$ Carinae consists of two distinct parts:
the Homunculus and the outer ejecta. The outer ejecta are mainly
a collection of numerous  filaments, shaped irregularly and distributed
over  an area  of  1\arcmin\,$\times$\,1\arcmin.
While the Homunculus is mainly a reflection nebula, the outer ejecta
are an emission nebula. Kinematic analysis of 
the outer ejecta (as the Homunculus) show their 
bi-directional expansion. Radial velocities in the
outer ejecta reach up to $>$2000\,\kms\ and the gas gives
rise to X-ray emission. 
The temperature of the  X-ray gas is of the order of
0.65 keV. 
These  shock temperatures 
indicate  velocities of the shocking gas of 750\,\kms,
about what was found for the average expansion velocity 
of the outer ejecta. 
HST/STIS data from the strings,  long, highly collimated structures in the
outer ejecta, show that the electron
density of the strings is of the order of 10$^4$\,cm$^{-3}$. 
Other structures in the outer ejecta show similar  values.
String 1 has a mass of about 3 10$^{-4}$\,M$_{\sun}$,
a  density gradient along the strings or a denser leading head
was not found. 
\end{abstract}

\section{The outer ejecta: Morphology and  Kinematics}

\begin{figure}
\plotone{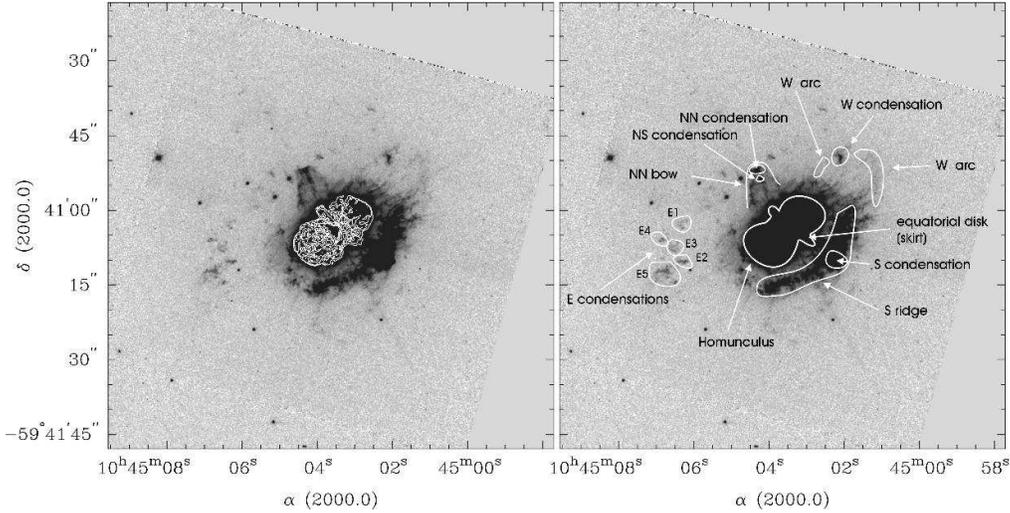}
\label{hst}
\caption{{\it Left:} An F658N HST image of  the Homunculus (additionally in
  contours) and  the outer ejecta. {\it Right:} This figure shows again
the HST images with the brightest structures  outlined and their 
designation given. Most parts of  the nomenclature are 
according to Walborn (1976). }
\end{figure}

$\eta$ Carinae, one of the most massive stars known,  is situated 
in the large H\,{\sc ii} region NGC 3372, and  is long known to be an 
irregularly variable source (Humphreys et al.\ 1999 and references therein). 
The most sudden change was  around  1843 when it brightened by  several
magnitudes to $-$1$^{\rm m}$. The  event is now 
known as the ``Great Eruption''. Shortly after the optical brightness 
decreased by more than  7$^{\rm m}$.  During this event a larger amount of mass
has been ejected and formed $\eta$ Carinae's circumstellar nebula.
The first images of this nebula were made by
Gaviola (1946) and Thackeray (1949). Gaviola named the nebula
according to the geometry he identified at that time the {\it Homunculus}.
It had a size of somewhat larger than 10\arcsec.
Today's images show the  Homunculus is  symmetric---bipolar---and
larger. It also  turned out, that the Homunculus is only the central part 
of the nebula around $\eta$ Carinae (e.g.  Walborn 1976). In total the 
nebula extends
to a diameter of 60\arcsec\ (0.67\,pc). The central
bipolar structure, stilled called 
Homunculus, has a diameter (long axis) of about 18\arcsec\ (0.2\,pc).  
All outer filaments (outside  the Homunculus)
are combined into what is known as the
{\it  outer ejecta}. The sizes and morphology of structures in the outer
ejecta are manifold. Contrary to the Homunculus, the outer ejecta are not
symmetric, nor a coherent structure, but rather contain
numerous filaments, bullets or knots.
The different morphology of the
Homunculus and the outer ejecta as well as  the large difference in
brightness, is illustrated in Fig.\  \ref{hst}
Here an HST image shows  the Homunculus, additionally plotted with contours,
and the outer ejecta seen through an F658N  filter. The Homunculus and
the outer ejecta are very prominent in the [N\,{\sc ii}] line. This is the
result of the higher nitrogen abundance in the total nebula,  due to the
presence of CNO processed material.

Analysis about the expansion velocities of the Homunculus
are about as  old as the first images. Ringuelet (1958) 
obtained 300\,\kms\ using proper motions (using however a too small  distance
of  1.4\,kpc, newer distance estimates yield about 2.3\,kpc, so this velocity
transforms into 500\,\kms). Thackeray (1961) obtains values of 
roughly 650\,\kms. With better spectroscopy and HST astrometry 
it was shown that the Homunculus, as images suggest,
expands  bipolar, with about 650\,\kms (e.g.\ Davidson \& Humphreys 1997, 
Currie et al. 1996) and with the south-east lobe approaching and the north-west
lobe receding.

Measurements of the expansion of the  outer ejecta have been made
determining again the
proper  motion of the brightest structure,  e.g.\ the E condensations 
(e.g.\  Walborn \& Blanco 1988).
Fabry-Perot measurements and high-resolution echelle observations
followed and  showed  that 
the outer ejecta has on average similar expansion velocities
as the Homunculus (e.g. Meaburn et al. 1996, Weis et al. 2001a).
The majority of the structures expanding with $|v_{\rm exp}|$ = 600-700\,\kms.
A significant fraction of the filaments, however, move much faster,
reaching velocities as high as 2000\,\kms (Weis 2001a,b).
A few structures are as fast as 2500-3000\,\kms  
(Weis et al.\ 2001b,  Smith \& Morse 2004). 
Additionally the kinematics show the outer ejecta to be
a more ordered structure  than  expected  from 
morphology. In the south-eastern
region most filaments are blueshifted, 
while in the north-west the clumps are redhifted, 
see left panel in Fig. \ref{kinx} Comparing this movement with 
the expansion  of the Homunculus 
the outer ejecta is following a similar expansion  pattern.
The outer ejecta expands  bi-directional and with about the
same symmetry axis as the Homunculus.

\section{The outer ejecta's X-ray emission}

\begin{figure}
\plotone{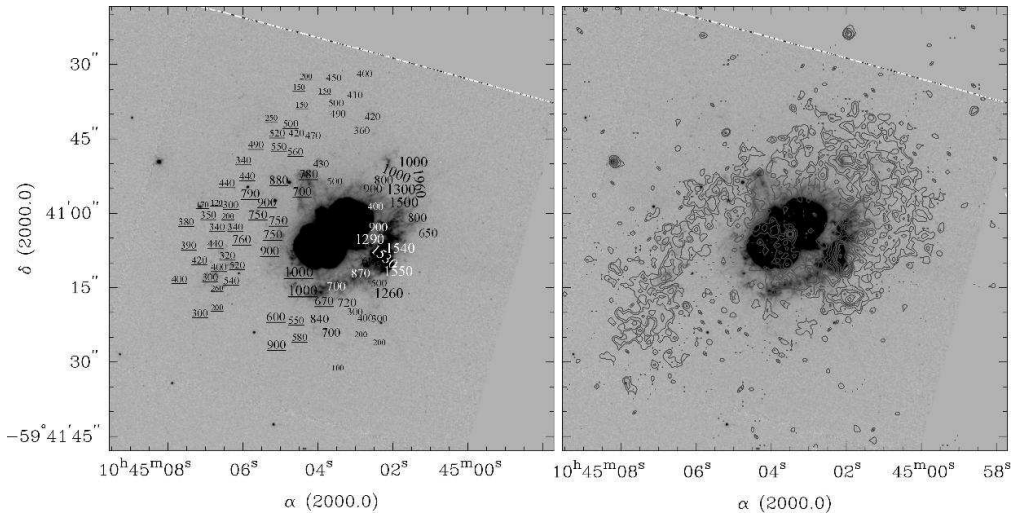}
\label{kinx}
\caption{{\it Left:} HST image of $\eta$ Carinae's outer ejecta, with 
radial expansion velocities overplotted.  Negative velocities
are underlined.  {\it Right:} Optical image and  in contours
the X-ray emission (0.6-1.2\,keV)  as detected with CHANDRA. 
}
\end{figure}

The nebula around $\eta$ Carinae is
detected in X-rays, as is the central object. Originally barely resolved with
the EINSTEIN satellite (Chlebowski et al.\ 1984) the
higher spatial resolution of ROSAT and CHANDRA makes it possible to better
separate the emission of the nebula from that of the harder central source.
Already with ROSAT it was found that the softer emission is roughly
hook shaped (e.g. Corcoran et al. 1994, Weis
et al.\ 2001a) and agrees in its dimensions with the extension of the
outer ejecta (Weis et al.\ 2001a).
The CHANDRA/ACIS data (Weis et al.\ 2004)  
are so far the images with the highest spatial
resolution and at the same time had decent spectral resolution. 
Fig. \ref{kinx} shows in contours the X-ray emission overplotted on an 
HST image.
The central source---which might be slightly extended---is very hard and
as known from other observations variable (e.g. Corcoran et al.\ 1995,
Ishibashi et al.\ 1999, Weis et al.\ 2001a). 
The emission
which results from the outer ejecta is much softer and
shows a hook shape, plus a bridge like connection which crosses the central
object from the south-west end of the hook to the ridge in the
north-east (Weis et al.\ 2004). 

The {\it S\,condensation} (Walborn 1976), is the brightest X-ray 
feature and agrees well with the bright optical filament.
The expansion velocities derived from our optical Echelle spectra
(see Weis 2001a,b) are plotted in Fig. \ref{kinx} (left).
In regions with on  average faster  moving clumps
the intensity of the  X-ray emission is higher, too.
The morphology of the extended soft X-ray emission
from $\eta$ Carinae can therefore be explained 
by shocks formed through the interaction of the
faster moving filaments.
Spectra extracted from the 0th order image of the
 CHANDRA/ACIS-S data have been modeled
using a one temperature, thermal equilibrium model (Mewe-Kaastra-Liedahl)
and a fixed column density.
Fits to the data yield an average temperature of $\sim$ 0.65 keV
(Weis et al. 2004). Gas of this  temperature 
indicates  velocities of the shocking gas
around 750\,\kms, in good agreement to the detected average expansion
velocities.

\section{The Strings in the outer ejecta}

\begin{figure}
\plotone{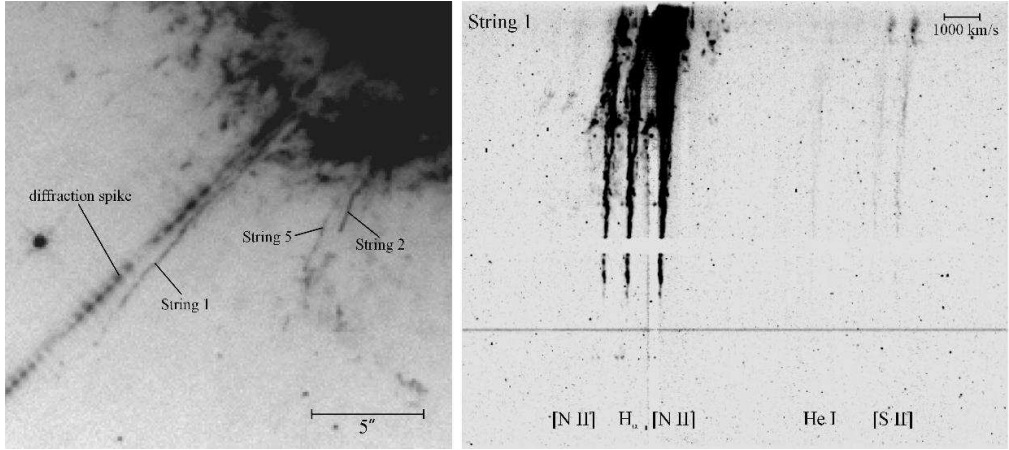}
\label{strings}
\caption{{\it Left:} Image of three strings in $\eta$ Carinae's outer ejecta.
{\it Right:} HST/STIS  spectrum of String 1.
}
\end{figure}

The {\it strings} are long, highly collimated structures that point nearly
radially away from $\eta$ Carinae and emerge from the outer ejecta
region that is closest to the Homunculus, partly that is the {\it S\,ridge}.
String\,1 (Fig.\ \ref{strings}) is the longest with
a total length 0.177\,pc and a width of equal or less than 0.003\,pc.
String\,4 (0.13\,pc), String\,3 (0.085\,pc), String\,5 (0.058\,pc), and
String\,2 (0.044\,pc) follow, all with similar
width. For a detailed description of the basic parameters of the strings see
Weis et al.\ (1999). The strings are nearly aligned with the long axis of the
Homunculus and their radial velocity increases as the string move outwards.
For String\,1 this increase ranges from $-522$ to 
about $-995$\,\kms (Weis et al.\ 1999). 
We observed the strings with the HST/STIS (G750M grating; 52\,$\times$\,0.5 
aperture)  with the slits aligned along the strings. 
The spectra ranged from 6295 to 6867\,\AA\ all strings
are identified in emission in the following lines: [N\,{\sc ii}] at 6548\,\AA\
and 6583\,\AA, H$_{\alpha}$, He\,{\sc i} at 6678\,\AA\ and both [S\,{\sc ii}]
 lines at 6716 and 6731\,\AA. Figure \ref{strings} shows String\,1 in these
 lines. 

Bullets originally believed to be the head of String\,1 are
 visible and appear detached from the string---below the continuum of the
 star in Fig. \ref{strings} (right)---they follow
the linear velocity increase well.
The spectra show  the strings to be more  a steady flow
rather than bullets along a chain.
The origin of String\,1 is sudden, within the southern part of the S ridge.
It seems not to extend back into the Homunculus, 
Fig.\,\ref{strings} (right).
Also a split which was suggested to occur in String\,1 (Weis et al.\ 1999)
was confirmed in the STIS data.
Using the [S\,{\sc ii}] line ratio and a temperature of about
14\,000\,K (a typical value in S ridge, Dufour et al.\ 1997)
we obtain an electron density of about 1.2 10$^4$ cm$^{-3}$. 
The ratio lies close to the high density limit and 
is more of a conservative lower limit and might be higher. The results
agree well with electron densities in the outer ejecta (S\,ridge)
determined by Dufour et al. (1997) using the Si\,{\sc iii}] lines.
With the help of the electron density and kinematics of String\,1 one can
calculate the mass and determine the kinetic energy associated with the
string. With the measured electron density we derive 
a mass density $\rho$ of 1.6 10$^{-23}$\,kg\,cm$^{-3}$
(using cosmic abundances). 
Using a volume of the string of 3.7 10$^{49}$\,cm$^{3}$, the mass $M$
of String\,1 is
about 3 10$^{-4}$\,M$_{\sun}$ if it is completely filled.
With a lower limit on the average expansion velocity
of $v\sim$450\,\kms\ the total kinetic
energy $E=1/2\,M\,v^2$ is about 6 10$^{43}$\,ergs.
Or using the cross section $A$,
a kinetic Luminosity $L=1/2\,\rho\,v^2\,A\,v$ can be estimated 
which is about 1\,L$_{\sun}$.
Several mechanism have been proposed to explain the strings, like trails left
by a bullet (Weis et al.\ 1999, Redman et al.\ 2002, Poludnenko et al.\ 2003), 
shadowing effects
(Soker 2001) or a steady gas flow (Weis et al.\ 1999).
The new data show the strings to be dense (10$^4$\,cm$^{-3}$; 
but might be hollow)
but no density gradient was detected
along the string which could give rise to a denser leading head, ruling out
the bullet scenario.

\end{document}